\documentclass[%
 reprint,
superscriptaddress,
 amsmath,amssymb,
 aps,
]{revtex4-2}

\usepackage[dvipsnames]{xcolor}
\usepackage{graphicx}
\usepackage{dcolumn}
\usepackage{bm}

\begin{document}

\preprint{APS/123-QED}

\title{Effects of Magnetic Fields in Hot White Dwarfs}

\author{J. Peterson}
\affiliation{Department of Physics, Kent State University, Kent OH 44242 USA}


\author{V. Dexheimer}
\email{vdexheim@kent.edu}
\affiliation{Department of Physics, Kent State University, Kent OH 44242 USA}

\author{R. Negreiros}
\affiliation{Department of Physics, Universidade Federal Fluminense, Niteroi, Brazil}

\author{B. G. Castanheira}
\affiliation{Department of Physics, Baylor University, Waco, TX 76798, USA}


\date{\today}

\begin{abstract}
In this work, we study the effects of temperature on magnetic white dwarfs. We model their interior as a nuclei lattice surrounded by a relativistic free Fermi gas of electrons, accounting for effects from temperature, Landau levels and anomalous magnetic moment. We find that, at low densities (corresponding to the outer regions of star), both temperature and magnetic field effects play an important role in the calculation of microscopic thermodynamical quantities. To study macroscopic stellar structures within a general-relativistic approach, we solve numerically the coupled Einstein's-Maxwell's equations for fixed entropy per particle configurations and discuss how temperature affects stellar magnetic field profiles, masses and radii.
\end{abstract}

                   
                              
\maketitle


\section{Introduction}

While the core of most white dwarf stars are adequately modeled with one of the assumptions that the temperature or the magnetic field can be disregarded, some recent observations (Ref.~\cite{Landstreet2020,Ferrario2015,Kilic2019,Kawka2019}) have suggested that a few white dwarfs may require the inclusion of both temperature and magnetic field effects in the calculation of the matter equation of state. In that light, we examine for the first time the effects of including both temperature and magnetic field into the equation of state of white dwarfs. We followed the relativistic formalism developed in Ref.~\cite{Strickland:2012vu} to describe a finite-temperature free Fermi gas under the influence of magnetic fields, including both, the particle quantization into discrete orbits perpendicular to the local direction of the magnetic field \cite{LandauLifshitz1977} and the anomalous magnetic moment \cite{PhysRev.173.1210,PhysRev.176.1438}, which differentiates between different particle spin projections. Whereas in Ref.~\cite{Strickland:2012vu}, we focused on nucleons and energy scales relevant for magnetic neutron stars and particle collisions, here, we focus on electrons (embedded in a lattice of Carbon nuclei) and energy scales relevant for white dwarfs. 

The typical white dwarf effective temperature may vary from $3,000$ to $140,000$ K and the core temperature from $10^6$ to $10^7$ K \cite{2001PASP..113..409F}. The atmospheric magnetic field, identified using Zeeman effect, can vary from $10^3$ to $10^9$ G \cite{Kawka2019,Kepler2013}, with the central one being impossible to measure, but possible to estimate with works such as this one. According to Ref.~\cite{Kawka2007}, as many as $20$\% of white dwarfs have surface magnetic fields of $B=10^7$ G or larger and, according to Ref.~\cite{Kawka2019}, magnetic white dwarfs tend to be more massive, possibly indicating  accretion \cite{Nordhaus2011} or mergers \cite{Briggs2018}. However, the uncertainty in the mass determination in magnetic white dwarfs is larger, because the hydrogen lines used in the spectroscopic fits are modified due to the magnetic field \cite{Kuelebi2009}. Finally, white dwarfs are expected to have central mass density varying from $10^6$ to $10^8$ $g/cm^3$ \cite{2001PASP..113..409F}, although the magnetic field tends to decrease stellar central density, in a mechanism not too different from rotation \cite{Franzon:2015sya}.

In this work, we study microscopic and macroscopic properties of white dwarfs. Strong magnetic fields affect the structure of the space-time metric, as they are a source for the gravitational field through the Maxwell energy-momentum tensor. As a consequence, magnetized stars are anisotropic and require a general-relativity treatment beyond the solution of the widely used Tolman-Oppenheimer-Volkoff (TOV) equations \cite{Oppenheimer:1939ne,Tolman:1939jz}. In this work, we model magnetic stars by solving Einstein-Maxwell equations in a similar way to what was done in Refs.~\cite{Bonazzola:1993zz,Cardall:2000bs}. More precisely, we numerically solve the coupled Einstein-Maxwell's equation for a compact object with a non-vanishing electric current. Such system of equations is solved iteratively by expanding the fields in appropriate Green's functions \cite{Cook1992a}.

We have chosen to show stellar configurations that have central magnetic field strengths $\sim B=10^{13}$ G. Several works cite this number as the limit for white dwarf stability, including spherical and axisymmetric solutions of Einstein's equations \cite{Paret:2015dja}. This limit also coincides with the threshold extracted from the Virial theorem using simple assumptions (see detailed discussion in Ref.~\cite{Coelho:2013bba}). Note that larger limits were discussed earlier in the literature using Newtonian solutions \cite{Das:2012ai} and, more recently, using axisymmetric solutions of Einstein's equations in the context of a softer equation of state generated by the inclusion of pycnonuclear fusion reactions \cite{otoniel2017magnetized}.

In Ref.~\cite{Franzon:2016iai}, we studied the effects of strong magnetic fields, finite temperature, and rotation on neutron stars. But in this case, no magnetic field effects were accounted for in the equation of state, only in the numerical general-relativity code, and the focus was on how the magnetic field affected the diverse particle population of the neutron star, not the equation of state or stellar mass-radius relation. On the other hand, thermal effects of the equation of state for quark stars we studied using spherical solutions in Refs.~\cite{Dexheimer:2012mk,LopezFune:2019hkh}. In this work, we start by reviewing the microscopic formalism used in Section II, followed by the microscopic results in Section III, and the macroscopic formalism and results in section IV. In Section V, we draw our conclusions and final discussion.

\section{Microscopic Formalism}

We model the interior of white dwarfs by starting with a relativistic free Fermi gas of electrons and include effects of both temperature and magnetic fields. To incorporate low temperature (compared to the Fermi energy) effects numerically, we split the integrals for each thermodynamic quantity into three parts with respect to the Fermi momentum $k_z$ going from:
\begin{itemize}
\item 0 to $\sqrt{(\mu-T)^2-\bar{m}^2}$;
\item $\sqrt{(\mu-T)^2-\bar{m}^2}$ to $\sqrt{(\mu+T)^2-\bar{m}^2}$;
\item beyond $\sqrt{(\mu+T)^2-\bar{m}^2}$,
\end{itemize}
with $\mu$ being the electron chemical potential, $T$ the temperature, and $\bar{m}$ the electron mass modified by the magnetic field. Naturally, the splitting only occurs when  $\sqrt{(\mu-T)^2-\bar{m}^2}$ and/or $\sqrt{(\mu+T)^2-\bar{m}^2}>0$. This procedure is necessary, once the deviation from a step function distribution function occurs for low temperatures in a very small range of Fermi energies.

For the equation of state, we assume the magnetic field $B$ to be (locally) pointing in the z-direction, so the electrons acquire discrete orbits in the plane perpendicular to the magnetic field $B$ creating a quantization of the energy levels along the x- and y-directions (Landau quantization). In addition, when the anomalous magnetic moment (AMM) of the electrons is included, we allow an asymmetry between the up and down spins of the electrons. More details about the effects of magnetic field and AMM on the free gas of fermions at finite temperature can be found in Ref.~\cite{Strickland:2012vu} and references therein. These are the expressions for number density, energy density, parallel pressure (z-direction), perpendicular pressure (x- and y- directions) and entropy density of spin one half fermions with charge $q$, in our case, the electrons:
\begin{eqnarray}
n_e=\frac{|q|B}{2\pi^2}\sum_{s=\pm1}\sum_{n=0}^{\infty}\int_{0}^{\infty}dk_z[f_+(E,T,\mu)-f_-(E,T,\mu)],\nonumber\\
\label{1}
\end{eqnarray}
\begin{eqnarray}
\epsilon_e&=&\frac{|q|B}{2\pi^2}\sum_{s=\pm1}\sum_{n=0}^{\infty}\int_{0}^{\infty}dk_zE\times[f_+(E,T,\mu)\nonumber\\&+&f_-(E,T,\mu)],
\label{eq2}
\end{eqnarray}
\begin{eqnarray}
P_{\parallel e}&=&\frac{|q|B}{2\pi^2}\sum_{s=\pm1}\sum_{n=0}^{\infty}\int_{0}^{\infty}dk_z\frac{k_z^2}{E}\times[f_+(E,T,\mu)\nonumber\\&+&f_-(E,T,\mu)],
\label{eq3}
\end{eqnarray}
\begin{eqnarray}
P_{\perp e}&=&\frac{|q|B^2}{2\pi^2}\sum_{s=\pm1}\sum_{n=0}^{\infty}\bar{m}(\nu)\left(\frac{|q|\nu}{\sqrt{m^2+2|q|B\nu}}-s\kappa\right)\nonumber\\&\times&\int_{0}^{\infty}dk_z\frac{1}{E}[f_+(E,T,\mu)+f_-(E,T,\mu)],
\label{eq4}
\end{eqnarray}
\begin{eqnarray}
s_e&=&\frac{|q|B}{2\pi^2}\sum_{s=\pm1}\sum_{n=0}^{\infty}\int_{0}^{\infty}dk_z\Bigg[(1-f_+)\ln\left(\frac{1}{1-f_+}\right)
\nonumber\\&+&f_+\ln\left(\frac{1}{f_+}\right)+(1-f_-)\ln\left(\frac{1}{1-f_-}\right)\nonumber\\&+&f_-\ln\left(\frac{1}{f_-}\right)\Bigg],
\label{eq5}
\end{eqnarray}
 where it is understood that the parallel and perpendicular pressures specify the components of the energy-momentum tensor $T^{\mu\nu}$ of matter in the local rest frame of the system. In addition, the magnetization of charged spin one half fermions, defined as the derivative of the grand-canonical potential, is Refs.~\cite{Strickland:2012vu, LandauLifshitzPitaevskii}
\begin{eqnarray}
\mathcal{M}_e=-\frac{\partial\Omega}{\partial B}=\frac{\partial P_{\parallel e}}{\partial B}=\frac{P_{\parallel e}-P_{\perp e}}{B}.
\label{eq6}
\end{eqnarray}

In the expressions above, we used that $\kappa=\kappa_i\mu_B=(1.16\times 10^{-3})(|q|/2m_e)$, with $\kappa_i$ being the coupling strength for the AMM and $\mu_B$ the Bohr magneton, the Fermi energy of particles $E=\sqrt{k_z^2+\bar{m}(\nu)^2}$, with $k_z$ being the particle momentum in the direction of the magnetic field and $\bar{m}(\nu)=\sqrt{m^2+2\nu|q|B}-s\kappa B$ the modified particle mass, and the Landau level $\nu=n+\frac{1}{2}-\frac{s}{2}\frac{q}{|q|}$, with n being the discretized orbital angular momentum of the particle in the transverse plane, and $s=\pm1$ the spin projection of the particle along the direction of the magnetic field. The distribution functions for particles $f_+$ and anti-particles $f_-$ are defined as:
\begin{eqnarray}
f_\pm(E,T,\mu)=\frac{1}{e^{(E\mp\mu)/T}+1}.
\label{eq7}
\end{eqnarray}

Compared to the degenerate electrons, the ions in the white-dwarf (in our case Carbon nuclei) only contribute significantly to the energy density of the system, given their high mass and low momentum. A detailed description of the equation of state of white dwarfs as well as the roles played by the ions and electron gas can be found in Ref. \cite{Shapiro:1983du}. Ignoring lattice effects the energy density of a white dwarf can be written as
\begin{eqnarray}
\epsilon_B=\frac{E_B}{V}=\frac{M_B}{A/n_B}=n_B\frac{M_B}{A}=2n_em_B,
\label{eq8}
\end{eqnarray}
where $V$ is the volume of the system, $E_B$ the energy of the nuclei approximated by their mass $M_B$, $A$ the number of baryons in the constituent nucleus (in our case Carbon, $A=12$), $n_B$ the number density of baryons (which is twice the electron number density $n_e$ due to charge neutrality and the isospin symmetry of carbon nuclei), and $m_B$ the mass of each baryon. The baryon (mass) density is defined as $\rho=m_B n_B$.

\section{Microscopic Results}

\begin{figure}[t!]
\vspace{3mm}
\includegraphics[trim={1.4cm 0 0 2.6cm},width=9.7cm]{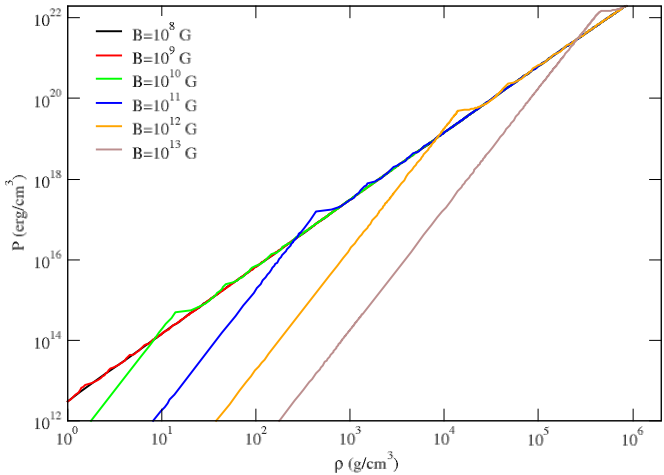}
\caption{(Color online) Parallel pressure as a function of baryon density for several magnetic field strengths assuming zero temperature ($T=0$).\label{0}}
\end{figure}

We start by showing the electron parallel pressure as a function of baryon density in the zero temperature case for several magnetic field strengths in Fig.~\ref{0}. Hereinafter, we refer to this pressure as $P$, as this is the thermodynamical quantity related to the grand-canonical potential density through $P_{\parallel e}=-\Omega$. First, we can immediately note the effect of Van Alphen oscillations (Ref.~\cite{VanHalfen}), whose oscillatory behavior is related to the discrete nature of the Landau levels; the first of these peaks occurs when the Landau level $\nu$ changes from zero to one, which happens at larger densities for larger magnetic fields. The AMM turns each peak into a double peak, though this is too subtle to be seen in the figures.

For the same density, there is a visible pattern that for $\nu=0$ the pressure is lower at higher magnetic fields until close to the first Van Alphen oscillation and as the magnetic field increases, it produces a slightly larger pressure beyond the first Van Alphen oscillation, where the oscillations are smaller and smoother due to the increasing number of Landau levels. In addition, for larger densities, it can be verified that the magnetic field makes the equation of state ($P_e$ vs. $\epsilon_e$) slightly stiffer. The AMM has no other overall significant effect, except at extremely lower densities, where it makes the equation of state a bit softer. This is due to the fact that only spin down electrons contribute to the zeroth Landau level, but they are suppressed due to the AMM positive coupling strength $\kappa_i$, decreasing the overall electron density. All results calculated with finite magnetic fields shown in this work include AMM effects, except when explicitly stated otherwise.

\begin{figure}[t!]
\vspace{3mm}
\includegraphics[trim={1.4cm 0 0 2.6cm},width=9.7cm]{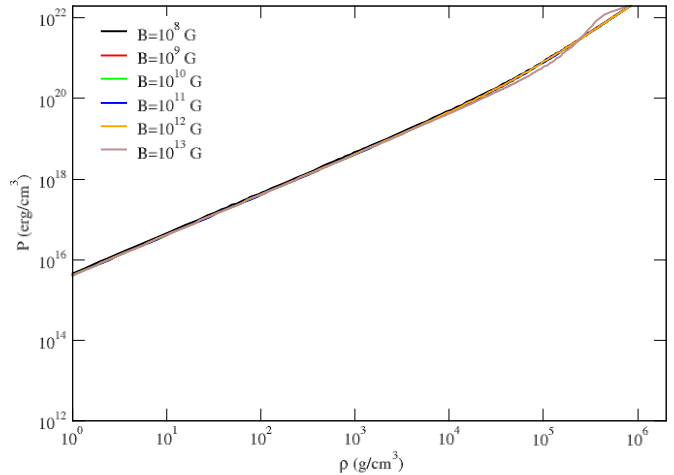}
\caption{(Color online) Parallel pressure as a function of baryon density for several magnetic field strengths assuming a fixed temperature of $T=10^8$ K.\label{2}}
\end{figure}

\begin{figure}[t!]
\vspace{3mm}
\includegraphics[trim={1.4cm 0 0 2.6cm},width=9.7cm]{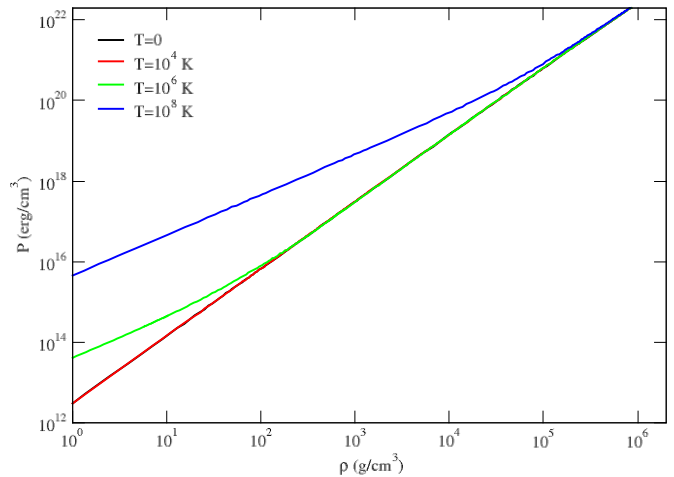}
\caption{(Color online) Parallel pressure as a function of baryon density for several temperatures assuming a constant magnetic field strength of $B=10^8$ G (which is effectively identical to the case with $B=0$).\label{3}}
\end{figure}

Fig.~\ref{2} also shows parallel pressure as a function of baryon density, except now with a large temperature $T=10^8$ K. The most notable difference is that the increased temperature diminishes the magnetic field effects and washes out the previously discussed Van Alphen oscillations, except for $B=10^{12}$ G and $B=10^{13}$ G.

\begin{figure}[t!]
\vspace{3mm}
\includegraphics[trim={1.4cm 0 0 2.6cm},width=9.7cm]{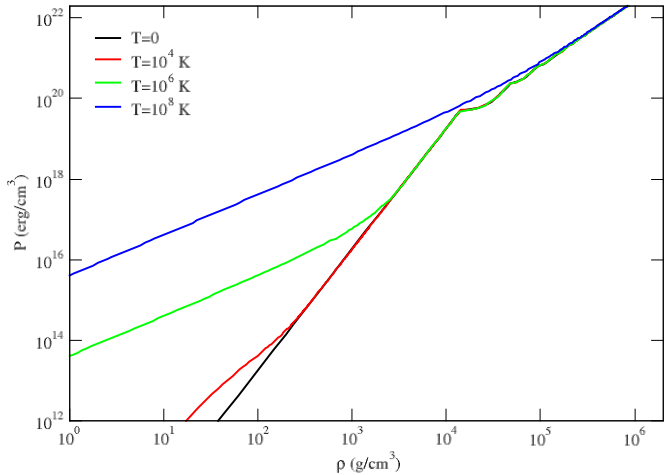}
\caption{(Color online) Parallel pressure as a function of baryon density for several temperatures assuming a constant magnetic field strength of $B=10^{12}$ G.\label{4}}
\end{figure}

As already shown in Ref.~\cite{2016IJMPS..4160129B}, in an approximation without antiparticles, temperature effects significantly increase the pressure (and the stiffness of the equation of state) of white dwarf matter at low densities. In Fig.~\ref{3}, we reproduce this feature for low (effectively zero) magnetic fields within our full approach including antiparticles. Note that for temperatures larger than $10^8$ K, the antiparticles would have a much more significant contribution, but this might not be relevant for the interior of white dwarfs. Fig.~\ref{4} shows that even for large magnetic fields, the temperature effects can still be observed at low densities with the additional effect of the Van Alphen oscillations.

\begin{figure}[t!]
\vspace{3mm}
\includegraphics[trim={1.4cm 0 0 2.6cm},width=9.7cm]{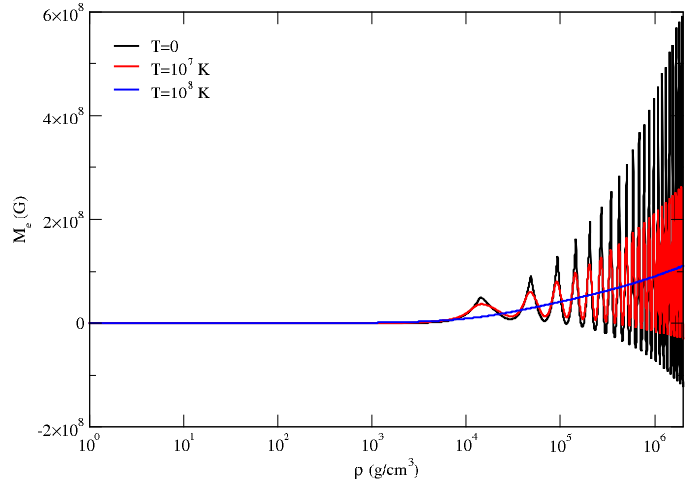}
\caption{(Color online) Magnetization as a function of baryon density for several temperatures assuming a constant magnetic field strength of $B=10^{12}$ G.\label{5}}
\end{figure}

\begin{figure}[t!]
\vspace{3mm}
\includegraphics[trim={1.4cm 0 0 2.6cm},width=9.7cm]{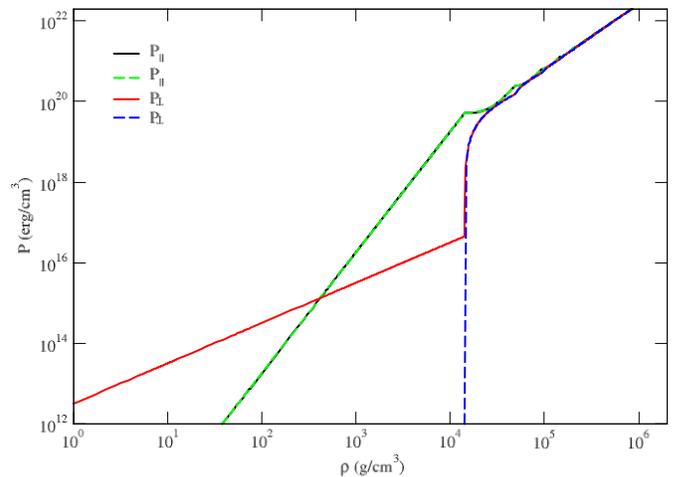}
\caption{(Color online) Both parallel and perpendicular pressures as a function of baryon density assuming zero temperature ($T=0$) and a constant magnetic field strength of $B=10^{12}$ G. The solid lines include anomalous magnetic moment (AMM) effects, while the dashed lines exclude it.\label{6}}
\end{figure}

Fig.~\ref{5} shows the magnetization of electrons as a function of density. The large oscillations at low temperature appear because the magnetization is the derivative of the parallel pressure (with respect to $B$), and therefore highlights small changes in pressure, enhancing the previously discussed Van Alphen oscillations. The oscillations wash out as the temperature increases and the effects from integer Landau levels become less important. The perpendicular pressure (shown in Fig.~\ref{6}) is small (different from zero only due to a small AMM correction) up to where the first non-zero Landau level appears. This can be seen in Eq.~(4). As a consequence, up to this point, the magnetization is simply $M=P_{||}/B$. After that, the magnetization oscillates going negative for baryon densities $n_B>~1.13\times10^5 g/cm^3$ (for the chosen magnetic field of $B=10^{12}$ G). Note that in Ref.~\cite{Strickland:2012vu} the enhanced AMM of the protons (second term in the parentheses of Eq.~(4) balanced the first term in the parentheses reducing the perpendicular pressure amount subtracted from the parallel pressure, resulting in a positive magnetization. In principle, the fact that the magnetization goes negative does not imply instability, since there are more contributions to the pressure, such as the pure magnetic field contribution (discussed in the next section), which is positive in the perpendicular direction.

Fig.~\ref{6} shows the parallel and perpendicular pressures; which are different as expected. As already mentioned, the perpendicular pressure is small up to the point where the first non-zero Landau level appears. This can be clearly seen in Fig.~\ref{6}. From Eq.~(6), it can be seen that the pressure in different directions differs by a factor that depends on the magnetization.

\begin{figure}[t!]
\vspace{3mm}
\includegraphics[trim={1.4cm 0 0 2.6cm},width=9.7cm]{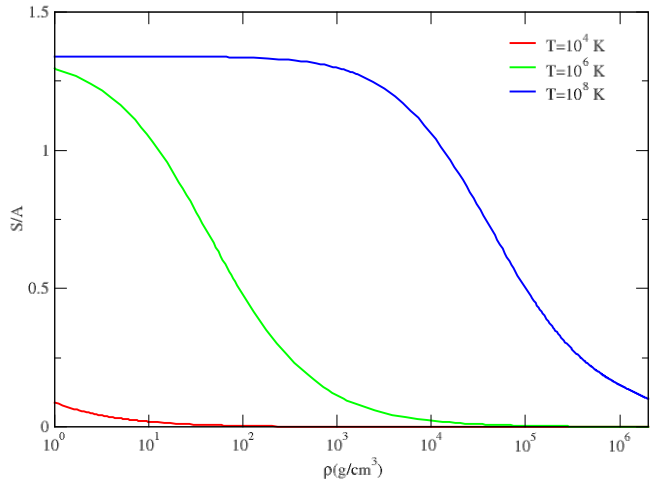}
\caption{(Color online) Entropy per baryon as a function of baryon density for several temperatures assuming there is no magnetic field ($B=0$).\label{7}}
\end{figure}

\begin{figure}[t!]
\vspace{3mm}
\includegraphics[trim={1.4cm 0cm. 0 4.77cm},width=9.7cm]{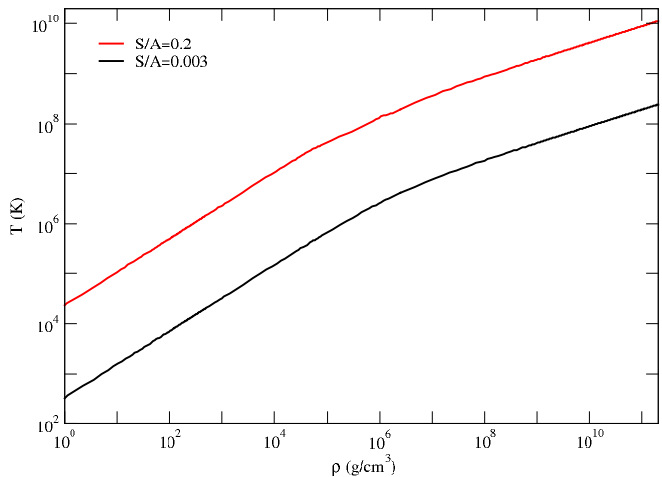}
\caption{(Color online) Temperature as a function of baryon density for two different values of entropy per baryon, assuming there is no magnetic field ($B=0$). The $\rho$ axis has been extended to show the full range covered in the next section.\label{8}}
\end{figure}

\begin{figure}[t!]
\vspace{3mm}
\includegraphics[trim={1.4cm 0 0 2.6cm},width=9.7cm]{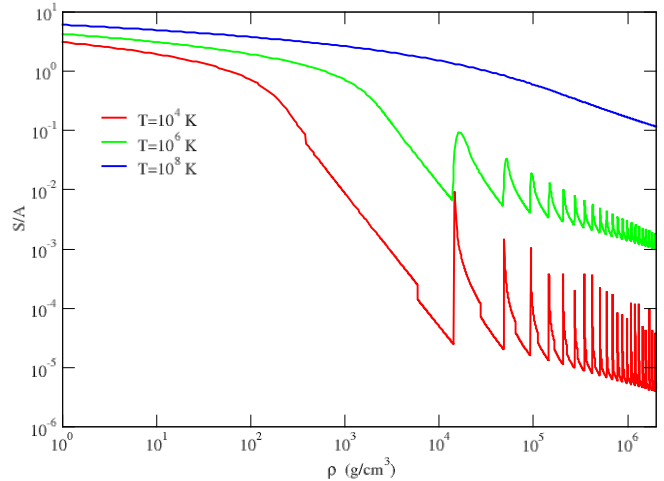}
\caption{(Color online) Entropy per baryon as a function of baryon density for several temperatures, assuming $B=10^{12}$ G.\label{9}}
\end{figure}

\begin{figure}[t!]
\vspace{3mm}
\includegraphics[trim={1.4cm 0 0 1.5cm},width=9.7cm]{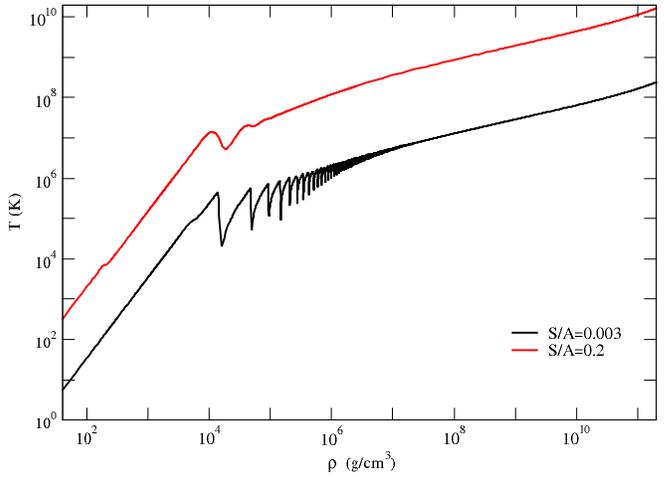}
\caption{(Color online) Temperature as a function of baryon density for two different values of entropy per baryon, assuming a constant magnetic field strength of $B=10^{12}$ G. The $\rho$ axis has been extended to show the full range covered in the next section.\label{10}}
\end{figure}

Fig.~\ref{7} shows the entropy per baryon as a function of baryon density for several fixed temperatures. For higher temperatures, the particles have higher entropy at any density. In compact stars, the temperature increases toward the center; a good way to simulate this effect is to fix the entropy per baryon and then calculate the temperature as a function of density. This is shown in Fig.~\ref{8} with two different values of entropy per baryon chosen in order to reproduce a realistic and more interesting scenario. Fig.~\ref{9} and Fig.~\ref{10} are the same as Fig.~\ref{7} and Fig.~\ref{8}, respectively, except now for a large magnetic field strength of $B=10^{12}$ G. The Van Alphen oscillations are prominent at lower temperatures. Note that the effect of anti-particles (properly accounted for in this work) becomes noticeable for some temperature between $10^8$ and $10^9$ K.

\section{Macroscopic Formalism and Results}

In order to obtain results for macroscopic stellar properties, such as mass and radius, in the case of magnetic stars, we have to simultaneously numerically solve the Einstein and Maxwell equations.
 We now briefly describe the structure equations for a general relativistic magnetized compact object. We begin by writing an axis-symmetric metric given by
\begin{eqnarray}\label{metric}
ds^2 =& -e^{2\nu}dt^2 + e^{2(\zeta - \nu)}(dr^2 + r^2d\theta ^2)  + \nonumber \\ &e^{-2\nu}G^2r^2sin^2\theta(d\phi - N^{\phi}dt)^2,
\end{eqnarray}
where the coordinates are  $x^{\mu}=(x^{0},x^{1},x^{2},x^{3})=(t,r,\theta,\phi)$ and the metric functions $\nu,\zeta, G$ and  $N^{\phi}$ depend on the coordinates $(r, \theta)$. The metric potentials are found by solving Einstein's equation coupled to Maxwell's equation in a curved space-time, these are respectively written as
\begin{eqnarray}\label{Einsteq}
G_{\mu \nu} = 8\pi T_{\mu \nu},
\end{eqnarray}
and
\begin{eqnarray}\label{maxwellinhomogeneous}
F^{\alpha \beta}_{\,\,\,\,\,\ ; \beta} = 4\pi j^{\alpha}.
\end{eqnarray}
where $G_{\mu \nu}$ is Einstein's tensor, $T_{\mu \nu}$ the energy-momentum tensor, $j^{\alpha}$ the four-current, and the Maxwell tensor $F_{\mu \nu}$ is given by
\begin{eqnarray}\label{Fmunu}
F_{\mu \nu} = A_{\nu, \mu} - A_{\mu, \nu},
\end{eqnarray}
where $A_\mu$ is the electromagnetic four potential. We note that comas and semi-colons have their usual meaning.

The matter-energy distribution is given by the energy-momentum tensor, which in this case is given by the sum of contributions coming from fermion matter that form a perfect fluid and that of the electromagnetic energy. We can therefore write
\begin{equation}\label{Tmunu}
T^{\mu \nu} = T^{PF \mu \nu} + T^{EM \mu \nu},
\end{equation}
where $T^{PF \mu \nu}$ denotes the perfect-fluid contribution, and $T^{EM \mu \nu}$ is the electromagnetic contribution. These are written as
\begin{eqnarray}\label{TPF}
T^{PF \mu \nu} = (\epsilon + P)u^{\mu}u^{\nu} + P g^{\mu \nu},
\end{eqnarray}
\begin{eqnarray}\label{TEM}
T^{EM \mu \nu} = \frac{1}{4\pi} \left( F^{\mu \alpha} F^{\nu}_{\,\,\, \alpha} - \frac{1}{4} g^{\mu \nu} F^{\alpha \beta} F_{\alpha \beta}  \right).
\end{eqnarray}
Finally, we need to ensure hydrostatic equilibrium in the star. The equilibrium condition may be obtained by the vanishing divergent of the energy-momentum tensor, which lead us to
\begin{eqnarray}\label{hydrostaticequil1}
\frac{1}{(\epsilon + P)} P_{,i} + \nu_{, i} - (\ln \Gamma)_{,i} - \frac{1}{(\epsilon + P)}f_{i} = 0,
\end{eqnarray}
where $\Gamma$ is the Lorentz factor and $f_i$ represents the Lorentz force and is given by
\begin{eqnarray}\label{lorenzforce}
f_{i} = F_{i \alpha} j^{\alpha} = j^{t}A_{t, i} + j^{\phi}A_{\phi, i}.
\end{eqnarray}

To fully define the problem we need to connect the macroscopic structure, defined by the above equations, to the miscroscopic realm. This is done via the equation of state ($P = P(\epsilon,T)$) and by defining a current function. In this work we adopt the following current function:
\begin{equation}
    j^\phi = f_0 (\epsilon + P),
\end{equation}
with $f_0$ being a current function that can be used to control the magnitude of the magnetic field. This choice of current leads to the formation of purely poloidal magnetic field, which is what we desire to study.

The set of equations described above is solved numerically by an iterative scheme that employs expansion in Green's functions. For a full description of the numerical technique employed, we refer the reader to \cite{Cook1992a}

\begin{figure}[t!]
\vspace{3mm}
\includegraphics[trim={1.4cm 0 0 2.6cm},width=9.7cm]{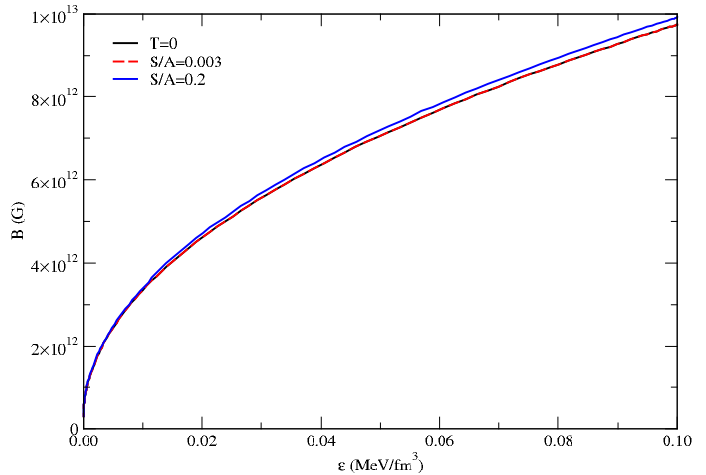}
\caption{(Color online) Magnetic field profile inside the most massive star of a sequence produced with current constant $f_0 = 10^{-3}$ in the polar direction as a function of energy density.\label{11}}
\end{figure}

It has been shown that including or not magnetic field effects in the equation of state when calculating certain macroscopic neutron star properties, such as mass, has little effect \cite{Chatterjee:2014qsa}. On the other hand, quantities related to the central stellar density that can change the particle population can be significant \cite{Franzon:2015sya}. In this work, as the particle population does not contain exotic matter (whose appearance depends on density), we use equations of state without magnetic field effects in the general relativity code to generate magnetic field profiles (as a function of energy density) and mass-radius relations.
In reality, the correct procedure would be to produce a 2-dimensional equation of state and allow the general relativity code to determine for a given central density and current function what is the magnetic field for a given energy density (as done in Refs.~\cite{Franzon:2015sya,Chatterjee:2014qsa} for neutron stars). However, as we restrict ourselves in this section to configurations that only reach a central magnetic field of $B\sim10^{13}$ G, less than the critical field $B_{\rm{crit}}=4.4\times10^{13}$ G in which the electron cyclotron energy equals its rest mass, this is not necessary.

\begin{figure}[t!]
\vspace{3mm}
\includegraphics[trim={1.4cm 0 0 2.6cm},width=9.7cm]{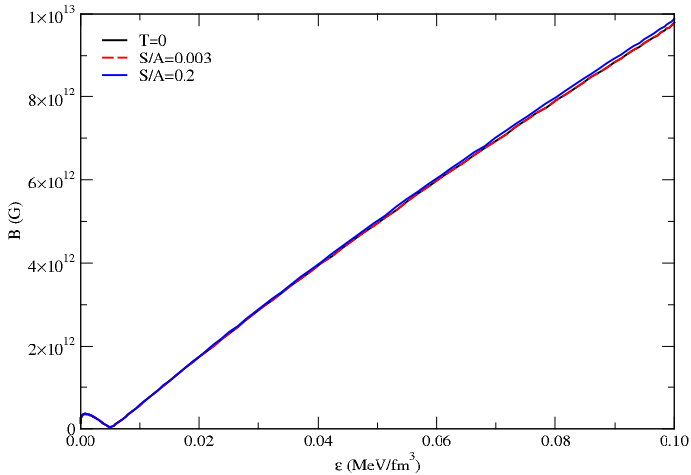}
\caption{(Color online) The same as in Fig.~11 but in the equatorial direction.\label{12}}
\end{figure}

Figs.~\ref{11} and \ref{12} show magnetic field profiles inside the most massive star of the sequence in the poloidal and equatorial directions. In Fig.~\ref{12} one can clearly see the effect of the Lorentz force reaching an extremum within the star rather than being monotonic; this manifests as the bump in the graph at low density. In both directions, the magnetic field strength is larger for a given energy density for larger entropies per particle (and temperature), although only the $S/A=0.2$ case presents a clearly visible difference at large temperatures. In this case, we see an increase of $1\%$ in the equatorial direction and $1.8\%$ in the poloidal direction at $\epsilon=0.1$ MeV/fm$^3$ and $T=10^{10}$ K. See Ref.~\cite{Gomes:2017zkc} for a detailed discussion between magnetic field strengths and equation of state stiffness in neutron stars with fixed currents. Here we have fixed the current constant $f_0$ to the value of $10^{-3}$.

\begin{figure}[t!]
\vspace{3mm}
\includegraphics[trim={1.4cm 0 0 2.6cm},width=9.7cm]{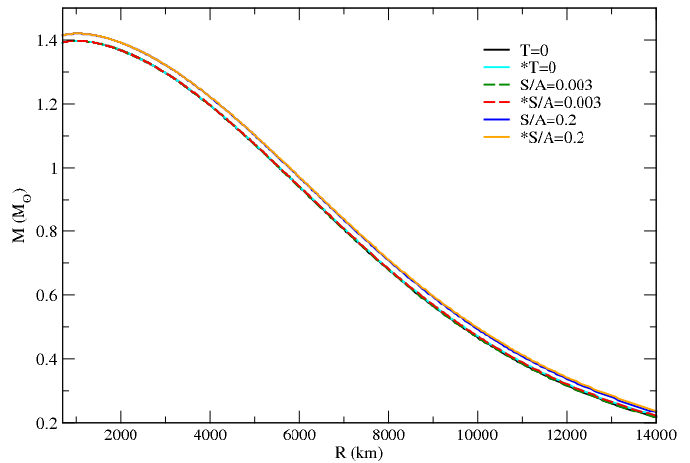}
\caption{(Color online) Mass-radius diagram for sequences of stars produced within several temperature scenarios. Lines marked with an asterisk denote sequences with magnetic field effects generated by fixing a current constant $f_0 = 10^{-3}$. \label{13}}
\end{figure}

Fig.~\ref{13} shows the mass-radius diagram for families of white dwarfs in different temperature scenarios, with and without magnetic fields effects and assuming a current constant $f_0 = 10^{-3}$ that generates the magnetic field profiles shown in Figs.~\ref{11} and \ref{12}. The figure shows that this strength of magnetic field is not enough to change stellar masses and radii. Nevertheless, the larger entropy per particle configuration generates, due to thermal effects, larger and more massive stars, increasing from a maximum mass of $1.397$ M$_\odot$ to $1.419$ M$_\odot$ with radii $\sim1000$ km.

Our results can be better put into context by looking at Fig.~9 of Ref.~\cite{Franzon:2015gda}, which shows that white dwarf masses are not modified by central magnetic fields  $B	\lesssim 10^{13}$ G, but seem to increase exponentially with magnetic fields beyond that. Note that, as shown in several figures of Ref.~\cite{Chatterjee:2016szk}, if we had used a Newtonian approach instead of ours, there would be an much larger (and nonphysical) mass change due to magnetic field effects.

\section{Conclusions and Discussion}

In this work, the equation of state for magnetic white dwarfs was modeled by using a finite temperature relativistic free Fermi gas of electrons embedded in a lattice of Carbon nuclei. We included effects coming from Landau level corrections, as well as from anomalous magnetic moment spin splitting. To our knowledge, this is the first time that simultaneous effects of including both temperature and magnetic field in the equation of state for white dwarfs was investigated. 

Focusing first on microscopic quantities, we saw that high temperatures tend to overpower the effects of magnetic fields expected to be seen in white dwarfs. At lower temperatures, the magnetic field effects are more pronounced with very visible Van Alphen oscillations. When looking at macroscopic quantities, the strong magnetic fields we considered were not large enough to change properties such as stellar masses and radii, although a finite temperature magnetic field profile in different directions of the star was extracted.

In the future, we intend on using these profiles to evaluate realistic magnetic field effects in, for example, pycnonuclear fusion reactions, and possibly on the crystalline structure of white dwarfs.

We would like to finalize by commenting on the realism of our choices. The choice of maximum magnetic field, $B=10^{13}$ G, is meant to examine the most extreme scenarios (similar to Ref.~\cite{Terrero2019}) but that may still be observed in white dwarfs. As for electron entropy per baryon, $S/A=0.003$ is in line with realistic upper bounds of white dwarf temperatures, topping out between $T=10^8$ K and $T=10^9$ K, the range in which a young white dwarf would exist as it was recently the core of a star undergoing the triple alpha process. $S/A=0.2$ tops out close to $T=10^{10}$ K and is meant to probe the extreme limits of white dwarfs.

The data presented in this work is available upon request and can be obtained by contacting the corresponding author.

\section*{Acknowledgements}

Support for this research comes from the National Science Foundation under grant PHY-1748621 and PHAROS (COST Action CA16214). R.N. acknowledges financial support from CAPES, CNPq and FAPERJ. This work is part of the project INCT-FNA Proc. No. 464898/2014-5 as well as FAPERJ JCNE Proc. No. E-26/203.299/2017.

\clearpage

\bibstyle{apsrev4-1}
\bibliography{apssamp}

\begin{thebibliography}{36}%
\makeatletter
\providecommand \@ifxundefined [1]{%
 \@ifx{#1\undefined}
}%
\providecommand \@ifnum [1]{%
 \ifnum #1\expandafter \@firstoftwo
 \else \expandafter \@secondoftwo
 \fi
}%
\providecommand \@ifx [1]{%
 \ifx #1\expandafter \@firstoftwo
 \else \expandafter \@secondoftwo
 \fi
}%
\providecommand \natexlab [1]{#1}%
\providecommand \enquote  [1]{``#1''}%
\providecommand \bibnamefont  [1]{#1}%
\providecommand \bibfnamefont [1]{#1}%
\providecommand \citenamefont [1]{#1}%
\providecommand \href@noop [0]{\@secondoftwo}%
\providecommand \href [0]{\begingroup \@sanitize@url \@href}%
\providecommand \@href[1]{\@@startlink{#1}\@@href}%
\providecommand \@@href[1]{\endgroup#1\@@endlink}%
\providecommand \@sanitize@url [0]{\catcode `\\12\catcode `\$12\catcode
  `\&12\catcode `\#12\catcode `\^12\catcode `\_12\catcode `\%12\relax}%
\providecommand \@@startlink[1]{}%
\providecommand \@@endlink[0]{}%
\providecommand \url  [0]{\begingroup\@sanitize@url \@url }%
\providecommand \@url [1]{\endgroup\@href {#1}{\urlprefix }}%
\providecommand \urlprefix  [0]{URL }%
\providecommand \Eprint [0]{\href }%
\providecommand \doibase [0]{https://doi.org/}%
\providecommand \selectlanguage [0]{\@gobble}%
\providecommand \bibinfo  [0]{\@secondoftwo}%
\providecommand \bibfield  [0]{\@secondoftwo}%
\providecommand \translation [1]{[#1]}%
\providecommand \BibitemOpen [0]{}%
\providecommand \bibitemStop [0]{}%
\providecommand \bibitemNoStop [0]{.\EOS\space}%
\providecommand \EOS [0]{\spacefactor3000\relax}%
\providecommand \BibitemShut  [1]{\csname bibitem#1\endcsname}%
\let\auto@bib@innerbib\@empty
\bibitem [{\citenamefont {Landstreet}\ and\ \citenamefont
  {Bagnulo}(2020)}]{Landstreet2020}%
  \BibitemOpen
  \bibfield  {author} {\bibinfo {author} {\bibfnamefont {J.~D.}\ \bibnamefont
  {Landstreet}}\ and\ \bibinfo {author} {\bibfnamefont {S.}~\bibnamefont
  {Bagnulo}},\ }\bibfield  {title} {\bibinfo {title} {Discovery of a
  sirius-like binary system with a very strongly magnetic white dwarf},\ }\href
  {https://doi.org/10.1051/0004-6361/201937301} {\bibfield  {journal} {\bibinfo
   {journal} {Astronomy and Astrophysics}\ }\textbf {\bibinfo {volume} {634}},\
  \bibinfo {pages} {L10} (\bibinfo {year} {2020})}\BibitemShut {NoStop}%
\bibitem [{\citenamefont {Ferrario}\ \emph {et~al.}(2015)\citenamefont
  {Ferrario}, \citenamefont {de~Martino},\ and\ \citenamefont
  {Gänsicke}}]{Ferrario2015}%
  \BibitemOpen
  \bibfield  {author} {\bibinfo {author} {\bibfnamefont {L.}~\bibnamefont
  {Ferrario}}, \bibinfo {author} {\bibfnamefont {D.}~\bibnamefont
  {de~Martino}},\ and\ \bibinfo {author} {\bibfnamefont {B.}~\bibnamefont
  {Gänsicke}},\ }\bibfield  {title} {\bibinfo {title} {Magnetic white
  dwarfs},\ }\href {https://doi.org/10.1007/s11214-015-0152-0} {\bibfield
  {journal} {\bibinfo  {journal} {Space Sci Rev}\ }\textbf {\bibinfo {volume}
  {191}},\ \bibinfo {pages} {111} (\bibinfo {year} {2015})}\BibitemShut
  {NoStop}%
\bibitem [{\citenamefont {Kilic}\ \emph {et~al.}(2019)\citenamefont {Kilic},
  \citenamefont {Rolland}, \citenamefont {Bergeron}, \citenamefont
  {Vanderbosch}, \citenamefont {Benni},\ and\ \citenamefont
  {Garlitz}}]{Kilic2019}%
  \BibitemOpen
  \bibfield  {author} {\bibinfo {author} {\bibfnamefont {M.}~\bibnamefont
  {Kilic}}, \bibinfo {author} {\bibfnamefont {B.}~\bibnamefont {Rolland}},
  \bibinfo {author} {\bibfnamefont {P.}~\bibnamefont {Bergeron}}, \bibinfo
  {author} {\bibfnamefont {Z.}~\bibnamefont {Vanderbosch}}, \bibinfo {author}
  {\bibfnamefont {P.}~\bibnamefont {Benni}},\ and\ \bibinfo {author}
  {\bibfnamefont {J.}~\bibnamefont {Garlitz}},\ }\bibfield  {title} {\bibinfo
  {title} {A magnetic white dwarf with five h-a components},\ }\href
  {https://doi.org/10.1093/mnras/stz2394} {\bibfield  {journal} {\bibinfo
  {journal} {Monthly Notices of the Royal Astronomical Society}\ }\textbf
  {\bibinfo {volume} {489}},\ \bibinfo {pages} {3648–3654} (\bibinfo {year}
  {2019})}\BibitemShut {NoStop}%
\bibitem [{\citenamefont {Kawka}(2019)}]{Kawka2019}%
  \BibitemOpen
  \bibfield  {author} {\bibinfo {author} {\bibfnamefont {A.}~\bibnamefont
  {Kawka}},\ }\bibfield  {title} {\bibinfo {title} {Clues to the origin and
  properties of magnetic white dwarfs},\ }\href
  {https://doi.org/10.1017/s1743921320000745} {\bibfield  {journal} {\bibinfo
  {journal} {Proceedings of the International Astronomical Union}\ }\textbf
  {\bibinfo {volume} {15}},\ \bibinfo {pages} {60–74} (\bibinfo {year}
  {2019})}\BibitemShut {NoStop}%
\bibitem [{\citenamefont {Strickland}\ \emph {et~al.}(2012)\citenamefont
  {Strickland}, \citenamefont {Dexheimer},\ and\ \citenamefont
  {Menezes}}]{Strickland:2012vu}%
  \BibitemOpen
  \bibfield  {author} {\bibinfo {author} {\bibfnamefont {M.}~\bibnamefont
  {Strickland}}, \bibinfo {author} {\bibfnamefont {V.}~\bibnamefont
  {Dexheimer}},\ and\ \bibinfo {author} {\bibfnamefont {D.~P.}\ \bibnamefont
  {Menezes}},\ }\bibfield  {title} {\bibinfo {title} {{Bulk Properties of a
  Fermi Gas in a Magnetic Field}},\ }\href
  {https://doi.org/10.1103/PhysRevD.86.125032} {\bibfield  {journal} {\bibinfo
  {journal} {Phys. Rev.}\ }\textbf {\bibinfo {volume} {D86}},\ \bibinfo {pages}
  {125032} (\bibinfo {year} {2012})},\ \Eprint
  {https://arxiv.org/abs/1209.3276} {arXiv:1209.3276 [nucl-th]} \BibitemShut
  {NoStop}%
\bibitem [{\citenamefont {Landau}\ and\ \citenamefont
  {Lifshitz}(1977)}]{LandauLifshitz1977}%
  \BibitemOpen
  \bibfield  {author} {\bibinfo {author} {\bibfnamefont {L.~D.}\ \bibnamefont
  {Landau}}\ and\ \bibinfo {author} {\bibfnamefont {E.~M.}\ \bibnamefont
  {Lifshitz}},\ }\href@noop {} {\emph {\bibinfo {title} {Quantum Mechanics:
  Non-Relativistic Theory}}}\ (\bibinfo  {publisher} {Pergamon Press},\
  \bibinfo {year} {1977})\BibitemShut {NoStop}%
\bibitem [{\citenamefont {Canuto}\ and\ \citenamefont
  {Chiu}(1968)}]{PhysRev.173.1210}%
  \BibitemOpen
  \bibfield  {author} {\bibinfo {author} {\bibfnamefont {V.}~\bibnamefont
  {Canuto}}\ and\ \bibinfo {author} {\bibfnamefont {H.-Y.}\ \bibnamefont
  {Chiu}},\ }\bibfield  {title} {\bibinfo {title} {Quantum theory of an
  electron gas in intense magnetic fields},\ }\href
  {https://doi.org/10.1103/PhysRev.173.1210} {\bibfield  {journal} {\bibinfo
  {journal} {Phys. Rev.}\ }\textbf {\bibinfo {volume} {173}},\ \bibinfo {pages}
  {1210} (\bibinfo {year} {1968})}\BibitemShut {NoStop}%
\bibitem [{\citenamefont {Chiu}\ \emph {et~al.}(1968)\citenamefont {Chiu},
  \citenamefont {Canuto},\ and\ \citenamefont
  {Fassio-Canuto}}]{PhysRev.176.1438}%
  \BibitemOpen
  \bibfield  {author} {\bibinfo {author} {\bibfnamefont {H.-Y.}\ \bibnamefont
  {Chiu}}, \bibinfo {author} {\bibfnamefont {V.}~\bibnamefont {Canuto}},\ and\
  \bibinfo {author} {\bibfnamefont {L.}~\bibnamefont {Fassio-Canuto}},\
  }\bibfield  {title} {\bibinfo {title} {Quantum theory of an electron gas with
  anomalous magnetic moments in intense magnetic fields},\ }\href
  {https://doi.org/10.1103/PhysRev.176.1438} {\bibfield  {journal} {\bibinfo
  {journal} {Phys. Rev.}\ }\textbf {\bibinfo {volume} {176}},\ \bibinfo {pages}
  {1438} (\bibinfo {year} {1968})}\BibitemShut {NoStop}%
\bibitem [{\citenamefont {{Fontaine}}\ \emph {et~al.}(2001)\citenamefont
  {{Fontaine}}, \citenamefont {{Brassard}},\ and\ \citenamefont
  {{Bergeron}}}]{2001PASP..113..409F}%
  \BibitemOpen
  \bibfield  {author} {\bibinfo {author} {\bibfnamefont {G.}~\bibnamefont
  {{Fontaine}}}, \bibinfo {author} {\bibfnamefont {P.}~\bibnamefont
  {{Brassard}}},\ and\ \bibinfo {author} {\bibfnamefont {P.}~\bibnamefont
  {{Bergeron}}},\ }\bibfield  {title} {\bibinfo {title} {{The Potential of
  White Dwarf Cosmochronology}},\ }\href {https://doi.org/10.1086/319535}
  {\bibfield  {journal} {\bibinfo  {journal} {Publications of the Astronomical
  Society of the Pacific}\ }\textbf {\bibinfo {volume} {113}},\ \bibinfo
  {pages} {409} (\bibinfo {year} {2001})}\BibitemShut {NoStop}%
\bibitem [{\citenamefont {{Kepler}}\ \emph {et~al.}(2013)\citenamefont
  {{Kepler}}, \citenamefont {{Pelisoli}}, \citenamefont {{Jordan}},
  \citenamefont {{Kleinman}}, \citenamefont {{Koester}}, \citenamefont
  {{K{\"u}lebi}}, \citenamefont {{Pe{\c{c}}anha}}, \citenamefont
  {{Castanheira}}, \citenamefont {{Nitta}}, \citenamefont {{Costa}},
  \citenamefont {{Winget}}, \citenamefont {{Kanaan}},\ and\ \citenamefont
  {{Fraga}}}]{Kepler2013}%
  \BibitemOpen
  \bibfield  {author} {\bibinfo {author} {\bibfnamefont {S.~O.}\ \bibnamefont
  {{Kepler}}}, \bibinfo {author} {\bibfnamefont {I.}~\bibnamefont
  {{Pelisoli}}}, \bibinfo {author} {\bibfnamefont {S.}~\bibnamefont
  {{Jordan}}}, \bibinfo {author} {\bibfnamefont {S.~J.}\ \bibnamefont
  {{Kleinman}}}, \bibinfo {author} {\bibfnamefont {D.}~\bibnamefont
  {{Koester}}}, \bibinfo {author} {\bibfnamefont {B.}~\bibnamefont
  {{K{\"u}lebi}}}, \bibinfo {author} {\bibfnamefont {V.}~\bibnamefont
  {{Pe{\c{c}}anha}}}, \bibinfo {author} {\bibfnamefont {B.~G.}\ \bibnamefont
  {{Castanheira}}}, \bibinfo {author} {\bibfnamefont {A.}~\bibnamefont
  {{Nitta}}}, \bibinfo {author} {\bibfnamefont {J.~E.~S.}\ \bibnamefont
  {{Costa}}}, \bibinfo {author} {\bibfnamefont {D.~E.}\ \bibnamefont
  {{Winget}}}, \bibinfo {author} {\bibfnamefont {A.}~\bibnamefont {{Kanaan}}},\
  and\ \bibinfo {author} {\bibfnamefont {L.}~\bibnamefont {{Fraga}}},\
  }\bibfield  {title} {\bibinfo {title} {{Magnetic white dwarf stars in the
  Sloan Digital Sky Survey}},\ }\href {https://doi.org/10.1093/mnras/sts522}
  {\bibfield  {journal} {\bibinfo  {journal} {Mon. Not. Roy. Astron. Soc.}\
  }\textbf {\bibinfo {volume} {429}},\ \bibinfo {pages} {2934} (\bibinfo {year}
  {2013})},\ \Eprint {https://arxiv.org/abs/1211.5709} {arXiv:1211.5709
  [astro-ph.SR]} \BibitemShut {NoStop}%
\bibitem [{\citenamefont {Kawka}\ \emph {et~al.}(2007)\citenamefont {Kawka},
  \citenamefont {Vennes}, \citenamefont {Schmidt}, \citenamefont
  {Wickramasinghe},\ and\ \citenamefont {Koch}}]{Kawka2007}%
  \BibitemOpen
  \bibfield  {author} {\bibinfo {author} {\bibfnamefont {A.}~\bibnamefont
  {Kawka}}, \bibinfo {author} {\bibfnamefont {S.}~\bibnamefont {Vennes}},
  \bibinfo {author} {\bibfnamefont {G.~D.}\ \bibnamefont {Schmidt}}, \bibinfo
  {author} {\bibfnamefont {D.~T.}\ \bibnamefont {Wickramasinghe}},\ and\
  \bibinfo {author} {\bibfnamefont {R.}~\bibnamefont {Koch}},\ }\bibfield
  {title} {\bibinfo {title} {Spectropolarimetric survey of hydrogen‐rich
  white dwarf stars},\ }\href {https://doi.org/10.1086/509072} {\bibfield
  {journal} {\bibinfo  {journal} {The Astrophysical Journal}\ }\textbf
  {\bibinfo {volume} {654}},\ \bibinfo {pages} {499–520} (\bibinfo {year}
  {2007})}\BibitemShut {NoStop}%
\bibitem [{\citenamefont {Nordhaus}\ \emph {et~al.}(2011)\citenamefont
  {Nordhaus}, \citenamefont {Wellons}, \citenamefont {Spiegel}, \citenamefont
  {Metzger},\ and\ \citenamefont {Blackman}}]{Nordhaus2011}%
  \BibitemOpen
  \bibfield  {author} {\bibinfo {author} {\bibfnamefont {J.}~\bibnamefont
  {Nordhaus}}, \bibinfo {author} {\bibfnamefont {S.}~\bibnamefont {Wellons}},
  \bibinfo {author} {\bibfnamefont {D.~S.}\ \bibnamefont {Spiegel}}, \bibinfo
  {author} {\bibfnamefont {B.~D.}\ \bibnamefont {Metzger}},\ and\ \bibinfo
  {author} {\bibfnamefont {E.~G.}\ \bibnamefont {Blackman}},\ }\bibfield
  {title} {\bibinfo {title} {Formation of high-field magnetic white dwarfs from
  common envelopes},\ }\href {https://doi.org/10.1073/pnas.1015005108}
  {\bibfield  {journal} {\bibinfo  {journal} {Proceedings of the National
  Academy of Sciences}\ }\textbf {\bibinfo {volume} {108}},\ \bibinfo {pages}
  {3135–3140} (\bibinfo {year} {2011})}\BibitemShut {NoStop}%
\bibitem [{\citenamefont {Briggs}\ \emph {et~al.}(2018)\citenamefont {Briggs},
  \citenamefont {Ferrario}, \citenamefont {Tout},\ and\ \citenamefont
  {Wickramasinghe}}]{Briggs2018}%
  \BibitemOpen
  \bibfield  {author} {\bibinfo {author} {\bibfnamefont {G.~P.}\ \bibnamefont
  {Briggs}}, \bibinfo {author} {\bibfnamefont {L.}~\bibnamefont {Ferrario}},
  \bibinfo {author} {\bibfnamefont {C.~A.}\ \bibnamefont {Tout}},\ and\
  \bibinfo {author} {\bibfnamefont {D.~T.}\ \bibnamefont {Wickramasinghe}},\
  }\bibfield  {title} {\bibinfo {title} {Genesis of magnetic fields in isolated
  white dwarfs},\ }\href {https://doi.org/10.1093/mnras/sty1150} {\bibfield
  {journal} {\bibinfo  {journal} {Monthly Notices of the Royal Astronomical
  Society}\ }\textbf {\bibinfo {volume} {478}},\ \bibinfo {pages} {899–905}
  (\bibinfo {year} {2018})}\BibitemShut {NoStop}%
\bibitem [{\citenamefont {{K{\"u}lebi}}\ \emph {et~al.}(2009)\citenamefont
  {{K{\"u}lebi}}, \citenamefont {{Jordan}}, \citenamefont {{Euchner}},
  \citenamefont {{G{\"a}nsicke}},\ and\ \citenamefont
  {{Hirsch}}}]{Kuelebi2009}%
  \BibitemOpen
  \bibfield  {author} {\bibinfo {author} {\bibfnamefont {B.}~\bibnamefont
  {{K{\"u}lebi}}}, \bibinfo {author} {\bibfnamefont {S.}~\bibnamefont
  {{Jordan}}}, \bibinfo {author} {\bibfnamefont {F.}~\bibnamefont {{Euchner}}},
  \bibinfo {author} {\bibfnamefont {B.~T.}\ \bibnamefont {{G{\"a}nsicke}}},\
  and\ \bibinfo {author} {\bibfnamefont {H.}~\bibnamefont {{Hirsch}}},\
  }\bibfield  {title} {\bibinfo {title} {{Analysis of hydrogen-rich magnetic
  white dwarfs detected in the Sloan Digital Sky Survey}},\ }\href
  {https://doi.org/10.1051/0004-6361/200912570} {\bibfield  {journal} {\bibinfo
   {journal} {Astron. Astrophys.}\ }\textbf {\bibinfo {volume} {506}},\
  \bibinfo {pages} {1341} (\bibinfo {year} {2009})},\ \Eprint
  {https://arxiv.org/abs/0907.2372} {arXiv:0907.2372 [astro-ph.SR]}
  \BibitemShut {NoStop}%
\bibitem [{\citenamefont {Franzon}\ \emph
  {et~al.}(2016{\natexlab{a}})\citenamefont {Franzon}, \citenamefont
  {Dexheimer},\ and\ \citenamefont {Schramm}}]{Franzon:2015sya}%
  \BibitemOpen
  \bibfield  {author} {\bibinfo {author} {\bibfnamefont {B.}~\bibnamefont
  {Franzon}}, \bibinfo {author} {\bibfnamefont {V.}~\bibnamefont {Dexheimer}},\
  and\ \bibinfo {author} {\bibfnamefont {S.}~\bibnamefont {Schramm}},\
  }\bibfield  {title} {\bibinfo {title} {{A self-consistent study of magnetic
  field effects on hybrid stars}},\ }\href
  {https://doi.org/10.1093/mnras/stv2606} {\bibfield  {journal} {\bibinfo
  {journal} {Mon. Not. Roy. Astron. Soc.}\ }\textbf {\bibinfo {volume} {456}},\
  \bibinfo {pages} {2937} (\bibinfo {year} {2016}{\natexlab{a}})},\ \Eprint
  {https://arxiv.org/abs/1508.04431} {arXiv:1508.04431 [astro-ph.HE]}
  \BibitemShut {NoStop}%
\bibitem [{\citenamefont {Oppenheimer}\ and\ \citenamefont
  {Volkoff}(1939)}]{Oppenheimer:1939ne}%
  \BibitemOpen
  \bibfield  {author} {\bibinfo {author} {\bibfnamefont {J.~R.}\ \bibnamefont
  {Oppenheimer}}\ and\ \bibinfo {author} {\bibfnamefont {G.~M.}\ \bibnamefont
  {Volkoff}},\ }\bibfield  {title} {\bibinfo {title} {{On Massive neutron
  cores}},\ }\href {https://doi.org/10.1103/PhysRev.55.374} {\bibfield
  {journal} {\bibinfo  {journal} {Phys. Rev.}\ }\textbf {\bibinfo {volume}
  {55}},\ \bibinfo {pages} {374} (\bibinfo {year} {1939})}\BibitemShut
  {NoStop}%
\bibitem [{\citenamefont {Tolman}(1939)}]{Tolman:1939jz}%
  \BibitemOpen
  \bibfield  {author} {\bibinfo {author} {\bibfnamefont {R.~C.}\ \bibnamefont
  {Tolman}},\ }\bibfield  {title} {\bibinfo {title} {{Static solutions of
  Einstein's field equations for spheres of fluid}},\ }\href
  {https://doi.org/10.1103/PhysRev.55.364} {\bibfield  {journal} {\bibinfo
  {journal} {Phys. Rev.}\ }\textbf {\bibinfo {volume} {55}},\ \bibinfo {pages}
  {364} (\bibinfo {year} {1939})}\BibitemShut {NoStop}%
\bibitem [{\citenamefont {Bonazzola}\ \emph {et~al.}(1993)\citenamefont
  {Bonazzola}, \citenamefont {Gourgoulhon}, \citenamefont {Salgado},\ and\
  \citenamefont {Marck}}]{Bonazzola:1993zz}%
  \BibitemOpen
  \bibfield  {author} {\bibinfo {author} {\bibfnamefont {S.}~\bibnamefont
  {Bonazzola}}, \bibinfo {author} {\bibfnamefont {E.}~\bibnamefont
  {Gourgoulhon}}, \bibinfo {author} {\bibfnamefont {M.}~\bibnamefont
  {Salgado}},\ and\ \bibinfo {author} {\bibfnamefont {J.~A.}\ \bibnamefont
  {Marck}},\ }\bibfield  {title} {\bibinfo {title} {{Axisymmetric rotating
  relativistic bodies: A new numerical approach for 'exact' solutions}},\
  }\href@noop {} {\bibfield  {journal} {\bibinfo  {journal} {Astron.
  Astrophys.}\ }\textbf {\bibinfo {volume} {278}},\ \bibinfo {pages} {421}
  (\bibinfo {year} {1993})}\BibitemShut {NoStop}%
\bibitem [{\citenamefont {Cardall}\ \emph {et~al.}(2001)\citenamefont
  {Cardall}, \citenamefont {Prakash},\ and\ \citenamefont
  {Lattimer}}]{Cardall:2000bs}%
  \BibitemOpen
  \bibfield  {author} {\bibinfo {author} {\bibfnamefont {C.~Y.}\ \bibnamefont
  {Cardall}}, \bibinfo {author} {\bibfnamefont {M.}~\bibnamefont {Prakash}},\
  and\ \bibinfo {author} {\bibfnamefont {J.~M.}\ \bibnamefont {Lattimer}},\
  }\bibfield  {title} {\bibinfo {title} {{Effects of strong magnetic fields on
  neutron star structure}},\ }\href {https://doi.org/10.1086/321370} {\bibfield
   {journal} {\bibinfo  {journal} {Astrophys. J.}\ }\textbf {\bibinfo {volume}
  {554}},\ \bibinfo {pages} {322} (\bibinfo {year} {2001})},\ \Eprint
  {https://arxiv.org/abs/astro-ph/0011148} {arXiv:astro-ph/0011148}
  \BibitemShut {NoStop}%
\bibitem [{\citenamefont {Cook}\ \emph {et~al.}(1992)\citenamefont {Cook},
  \citenamefont {Shapiro},\ and\ \citenamefont {Teukolsky}}]{Cook1992a}%
  \BibitemOpen
  \bibfield  {author} {\bibinfo {author} {\bibfnamefont {G.~B.}\ \bibnamefont
  {Cook}}, \bibinfo {author} {\bibfnamefont {S.~L.}\ \bibnamefont {Shapiro}},\
  and\ \bibinfo {author} {\bibfnamefont {S.~A.}\ \bibnamefont {Teukolsky}},\
  }\bibfield  {title} {\bibinfo {title} {{Spin-up of a rapidly rotating star by
  angular momentum loss - Effects of general relativity}},\ }\href
  {https://doi.org/10.1086/171849} {\bibfield  {journal} {\bibinfo  {journal}
  {The Astrophysical Journal}\ }\textbf {\bibinfo {volume} {398}},\ \bibinfo
  {pages} {203} (\bibinfo {year} {1992})},\ \Eprint
  {https://arxiv.org/abs/arXiv:1011.1669v3} {arXiv:arXiv:1011.1669v3}
  \BibitemShut {NoStop}%
\bibitem [{\citenamefont {Paret}\ \emph {et~al.}(2015)\citenamefont {Paret},
  \citenamefont {Horvath},\ and\ \citenamefont
  {Mart\'\i{}nez}}]{Paret:2015dja}%
  \BibitemOpen
  \bibfield  {author} {\bibinfo {author} {\bibfnamefont {D.~M.}\ \bibnamefont
  {Paret}}, \bibinfo {author} {\bibfnamefont {J.~E.}\ \bibnamefont {Horvath}},\
  and\ \bibinfo {author} {\bibfnamefont {A.~P.}\ \bibnamefont
  {Mart\'\i{}nez}},\ }\bibfield  {title} {\bibinfo {title} {{Maximum mass of
  magnetic white dwarfs}},\ }\href
  {https://doi.org/10.1088/1674-4527/15/10/010} {\bibfield  {journal} {\bibinfo
   {journal} {Res. Astron. Astrophys.}\ }\textbf {\bibinfo {volume} {15}},\
  \bibinfo {pages} {1735} (\bibinfo {year} {2015})},\ \Eprint
  {https://arxiv.org/abs/1501.04619} {arXiv:1501.04619 [astro-ph.HE]}
  \BibitemShut {NoStop}%
\bibitem [{\citenamefont {Coelho}\ \emph {et~al.}(2014)\citenamefont {Coelho},
  \citenamefont {Marinho}, \citenamefont {Malheiro}, \citenamefont {Negreiros},
  \citenamefont {Rueda}, \citenamefont {Ruffini},\ and\ \citenamefont
  {C\'aceres}}]{Coelho:2013bba}%
  \BibitemOpen
  \bibfield  {author} {\bibinfo {author} {\bibfnamefont {J.~G.}\ \bibnamefont
  {Coelho}}, \bibinfo {author} {\bibfnamefont {R.~M.}\ \bibnamefont {Marinho}},
  \bibinfo {author} {\bibfnamefont {M.}~\bibnamefont {Malheiro}}, \bibinfo
  {author} {\bibfnamefont {R.}~\bibnamefont {Negreiros}}, \bibinfo {author}
  {\bibfnamefont {J.~A.}\ \bibnamefont {Rueda}}, \bibinfo {author}
  {\bibfnamefont {R.}~\bibnamefont {Ruffini}},\ and\ \bibinfo {author}
  {\bibfnamefont {D.~L.}\ \bibnamefont {C\'aceres}},\ }\bibfield  {title}
  {\bibinfo {title} {{Dynamical instability of white dwarfs and breaking of
  spherical symmetry under the presence of extreme magnetic fields}},\ }\href
  {https://doi.org/10.1088/0004-637X/794/1/86} {\bibfield  {journal} {\bibinfo
  {journal} {Astrophys. J.}\ }\textbf {\bibinfo {volume} {794}},\ \bibinfo
  {pages} {86} (\bibinfo {year} {2014})},\ \Eprint
  {https://arxiv.org/abs/1306.4658} {arXiv:1306.4658 [astro-ph.SR]}
  \BibitemShut {NoStop}%
\bibitem [{\citenamefont {Das}\ and\ \citenamefont
  {Mukhopadhyay}(2012)}]{Das:2012ai}%
  \BibitemOpen
  \bibfield  {author} {\bibinfo {author} {\bibfnamefont {U.}~\bibnamefont
  {Das}}\ and\ \bibinfo {author} {\bibfnamefont {B.}~\bibnamefont
  {Mukhopadhyay}},\ }\bibfield  {title} {\bibinfo {title} {{Strongly magnetized
  cold electron degenerate gas: Mass-radius relation of the magnetized white
  dwarf}},\ }\href {https://doi.org/10.1103/PhysRevD.86.042001} {\bibfield
  {journal} {\bibinfo  {journal} {Phys. Rev. D}\ }\textbf {\bibinfo {volume}
  {86}},\ \bibinfo {pages} {042001} (\bibinfo {year} {2012})},\ \Eprint
  {https://arxiv.org/abs/1204.1262} {arXiv:1204.1262 [astro-ph.HE]}
  \BibitemShut {NoStop}%
\bibitem [{\citenamefont {Otoniel}\ \emph {et~al.}(2017)\citenamefont
  {Otoniel}, \citenamefont {Franzon}, \citenamefont {Malheiro}, \citenamefont
  {Schramm},\ and\ \citenamefont {Weber}}]{otoniel2017magnetized}%
  \BibitemOpen
  \bibfield  {author} {\bibinfo {author} {\bibfnamefont {E.}~\bibnamefont
  {Otoniel}}, \bibinfo {author} {\bibfnamefont {B.}~\bibnamefont {Franzon}},
  \bibinfo {author} {\bibfnamefont {M.}~\bibnamefont {Malheiro}}, \bibinfo
  {author} {\bibfnamefont {S.}~\bibnamefont {Schramm}},\ and\ \bibinfo {author}
  {\bibfnamefont {F.}~\bibnamefont {Weber}},\ }\href@noop {} {\bibinfo {title}
  {Very magnetized white dwarfs with axisymmetric magnetic field and the
  importance of the electron capture and pycnonuclear fusion reactions for
  their stability}} (\bibinfo {year} {2017}),\ \Eprint
  {https://arxiv.org/abs/1609.05994} {arXiv:1609.05994 [astro-ph.SR]}
  \BibitemShut {NoStop}%
\bibitem [{\citenamefont {Franzon}\ \emph
  {et~al.}(2016{\natexlab{b}})\citenamefont {Franzon}, \citenamefont
  {Dexheimer},\ and\ \citenamefont {Schramm}}]{Franzon:2016iai}%
  \BibitemOpen
  \bibfield  {author} {\bibinfo {author} {\bibfnamefont {B.}~\bibnamefont
  {Franzon}}, \bibinfo {author} {\bibfnamefont {V.}~\bibnamefont {Dexheimer}},\
  and\ \bibinfo {author} {\bibfnamefont {S.}~\bibnamefont {Schramm}},\
  }\bibfield  {title} {\bibinfo {title} {{Internal composition of proto-neutron
  stars under strong magnetic fields}},\ }\href
  {https://doi.org/10.1103/PhysRevD.94.044018} {\bibfield  {journal} {\bibinfo
  {journal} {Phys. Rev. D}\ }\textbf {\bibinfo {volume} {94}},\ \bibinfo
  {pages} {044018} (\bibinfo {year} {2016}{\natexlab{b}})},\ \Eprint
  {https://arxiv.org/abs/1606.04843} {arXiv:1606.04843 [astro-ph.HE]}
  \BibitemShut {NoStop}%
\bibitem [{\citenamefont {Dexheimer}\ \emph {et~al.}(2014)\citenamefont
  {Dexheimer}, \citenamefont {Menezes},\ and\ \citenamefont
  {Strickland}}]{Dexheimer:2012mk}%
  \BibitemOpen
  \bibfield  {author} {\bibinfo {author} {\bibfnamefont {V.}~\bibnamefont
  {Dexheimer}}, \bibinfo {author} {\bibfnamefont {D.~P.}\ \bibnamefont
  {Menezes}},\ and\ \bibinfo {author} {\bibfnamefont {M.}~\bibnamefont
  {Strickland}},\ }\bibfield  {title} {\bibinfo {title} {{The influence of
  strong magnetic fields on proto-quark stars}},\ }\href
  {https://doi.org/10.1088/0954-3899/41/1/015203} {\bibfield  {journal}
  {\bibinfo  {journal} {J. Phys. G}\ }\textbf {\bibinfo {volume} {41}},\
  \bibinfo {pages} {015203} (\bibinfo {year} {2014})},\ \Eprint
  {https://arxiv.org/abs/1210.4526} {arXiv:1210.4526 [nucl-th]} \BibitemShut
  {NoStop}%
\bibitem [{\citenamefont {L\'opez~Fune}(2019)}]{LopezFune:2019hkh}%
  \BibitemOpen
  \bibfield  {author} {\bibinfo {author} {\bibfnamefont {E.}~\bibnamefont
  {L\'opez~Fune}},\ }\href@noop {} {\bibinfo {title} {{Magnetized strange quark
  matter under stellar equilibrium and finite temperature}}} (\bibinfo {year}
  {2019}),\ \Eprint {https://arxiv.org/abs/1902.02717} {arXiv:1902.02717
  [astro-ph.HE]} \BibitemShut {NoStop}%
\bibitem [{\citenamefont {Landau}\ \emph {et~al.}(1984)\citenamefont {Landau},
  \citenamefont {Lifshitz},\ and\ \citenamefont
  {Pitaevskii}}]{LandauLifshitzPitaevskii}%
  \BibitemOpen
  \bibfield  {author} {\bibinfo {author} {\bibfnamefont {L.~D.}\ \bibnamefont
  {Landau}}, \bibinfo {author} {\bibfnamefont {E.~M.}\ \bibnamefont
  {Lifshitz}},\ and\ \bibinfo {author} {\bibfnamefont {L.~P.}\ \bibnamefont
  {Pitaevskii}},\ }\href@noop {} {\emph {\bibinfo {title} {Electrodynamics of
  Continuous Media. Vol. 8}}}\ (\bibinfo  {publisher} {Butterworth-Heinemann},\
  \bibinfo {year} {1984})\BibitemShut {NoStop}%
\bibitem [{\citenamefont {Shapiro}\ and\ \citenamefont
  {Teukolsky}(1983)}]{Shapiro:1983du}%
  \BibitemOpen
  \bibfield  {author} {\bibinfo {author} {\bibfnamefont {S.~L.}\ \bibnamefont
  {Shapiro}}\ and\ \bibinfo {author} {\bibfnamefont {S.~A.}\ \bibnamefont
  {Teukolsky}},\ }\href@noop {} {\emph {\bibinfo {title} {{Black holes, white
  dwarfs, and neutron stars: The physics of compact objects}}}}\ (\bibinfo
  {publisher} {Wiley},\ \bibinfo {year} {1983})\BibitemShut {NoStop}%
\bibitem [{\citenamefont {De~Haas}\ and\ \citenamefont
  {Van~Alphen}(1930)}]{VanHalfen}%
  \BibitemOpen
  \bibfield  {author} {\bibinfo {author} {\bibfnamefont {W.~J.}\ \bibnamefont
  {De~Haas}}\ and\ \bibinfo {author} {\bibfnamefont {P.~M.}\ \bibnamefont
  {Van~Alphen}},\ }\bibfield  {title} {\bibinfo {title} {The dependence of the
  susceptibility of diamagnetic metals upon the field - communication no. 212a
  from the physical labroratory, leiden},\ }\href
  {https://www.dwc.knaw.nl/DL/publications/PU00015989.pdf} {\bibfield
  {journal} {\bibinfo  {journal} {Proc. Acad. Sci. Amst.}\ }\textbf {\bibinfo
  {volume} {33}},\ \bibinfo {pages} {1106} (\bibinfo {year}
  {1930})}\BibitemShut {NoStop}%
\bibitem [{\citenamefont {{Boshkayev}}\ \emph {et~al.}(2016)\citenamefont
  {{Boshkayev}}, \citenamefont {{Rueda}}, \citenamefont {{Zhami}},
  \citenamefont {{Kalymova}},\ and\ \citenamefont
  {{Balgymbekov}}}]{2016IJMPS..4160129B}%
  \BibitemOpen
  \bibfield  {author} {\bibinfo {author} {\bibfnamefont {K.~A.}\ \bibnamefont
  {{Boshkayev}}}, \bibinfo {author} {\bibfnamefont {J.~A.}\ \bibnamefont
  {{Rueda}}}, \bibinfo {author} {\bibfnamefont {B.~A.}\ \bibnamefont
  {{Zhami}}}, \bibinfo {author} {\bibfnamefont {Z.~A.}\ \bibnamefont
  {{Kalymova}}},\ and\ \bibinfo {author} {\bibfnamefont {G.~S.}\ \bibnamefont
  {{Balgymbekov}}},\ }\bibfield  {title} {\bibinfo {title} {{Equilibrium
  structure of white dwarfs at finite temperatures}},\ }in\ \href
  {https://doi.org/10.1142/S2010194516601290} {\emph {\bibinfo {booktitle}
  {International Journal of Modern Physics Conference Series}}},\ \bibinfo
  {series} {International Journal of Modern Physics Conference Series},
  Vol.~\bibinfo {volume} {41}\ (\bibinfo {year} {2016})\ p.\ \bibinfo {pages}
  {1660129},\ \Eprint {https://arxiv.org/abs/1510.02024} {arXiv:1510.02024
  [astro-ph.SR]} \BibitemShut {NoStop}%
\bibitem [{\citenamefont {Chatterjee}\ \emph {et~al.}(2015)\citenamefont
  {Chatterjee}, \citenamefont {Elghozi}, \citenamefont {Novak},\ and\
  \citenamefont {Oertel}}]{Chatterjee:2014qsa}%
  \BibitemOpen
  \bibfield  {author} {\bibinfo {author} {\bibfnamefont {D.}~\bibnamefont
  {Chatterjee}}, \bibinfo {author} {\bibfnamefont {T.}~\bibnamefont {Elghozi}},
  \bibinfo {author} {\bibfnamefont {J.}~\bibnamefont {Novak}},\ and\ \bibinfo
  {author} {\bibfnamefont {M.}~\bibnamefont {Oertel}},\ }\bibfield  {title}
  {\bibinfo {title} {{Consistent neutron star models with magnetic field
  dependent equations of state}},\ }\href
  {https://doi.org/10.1093/mnras/stu2706} {\bibfield  {journal} {\bibinfo
  {journal} {Mon. Not. Roy. Astron. Soc.}\ }\textbf {\bibinfo {volume} {447}},\
  \bibinfo {pages} {3785} (\bibinfo {year} {2015})},\ \Eprint
  {https://arxiv.org/abs/1410.6332} {arXiv:1410.6332 [astro-ph.HE]}
  \BibitemShut {NoStop}%
\bibitem [{\citenamefont {Gomes}\ \emph {et~al.}(2017)\citenamefont {Gomes},
  \citenamefont {Franzon}, \citenamefont {Dexheimer},\ and\ \citenamefont
  {Schramm}}]{Gomes:2017zkc}%
  \BibitemOpen
  \bibfield  {author} {\bibinfo {author} {\bibfnamefont {R.~O.}\ \bibnamefont
  {Gomes}}, \bibinfo {author} {\bibfnamefont {B.}~\bibnamefont {Franzon}},
  \bibinfo {author} {\bibfnamefont {V.}~\bibnamefont {Dexheimer}},\ and\
  \bibinfo {author} {\bibfnamefont {S.}~\bibnamefont {Schramm}},\ }\bibfield
  {title} {\bibinfo {title} {{Many-body forces in magnetic neutron stars}},\
  }\href {https://doi.org/10.3847/1538-4357/aa8b68} {\bibfield  {journal}
  {\bibinfo  {journal} {Astrophys. J.}\ }\textbf {\bibinfo {volume} {850}},\
  \bibinfo {pages} {20} (\bibinfo {year} {2017})},\ \Eprint
  {https://arxiv.org/abs/1709.01017} {arXiv:1709.01017 [nucl-th]} \BibitemShut
  {NoStop}%
\bibitem [{\citenamefont {Franzon}\ and\ \citenamefont
  {Schramm}(2015)}]{Franzon:2015gda}%
  \BibitemOpen
  \bibfield  {author} {\bibinfo {author} {\bibfnamefont {B.}~\bibnamefont
  {Franzon}}\ and\ \bibinfo {author} {\bibfnamefont {S.}~\bibnamefont
  {Schramm}},\ }\bibfield  {title} {\bibinfo {title} {{Effects of strong
  magnetic fields and rotation on white dwarf structure}},\ }\href
  {https://doi.org/10.1103/PhysRevD.92.083006} {\bibfield  {journal} {\bibinfo
  {journal} {Phys. Rev. D}\ }\textbf {\bibinfo {volume} {92}},\ \bibinfo
  {pages} {083006} (\bibinfo {year} {2015})},\ \Eprint
  {https://arxiv.org/abs/1507.05557} {arXiv:1507.05557 [astro-ph.SR]}
  \BibitemShut {NoStop}%
\bibitem [{\citenamefont {Chatterjee}\ \emph {et~al.}(2017)\citenamefont
  {Chatterjee}, \citenamefont {Fantina}, \citenamefont {Chamel}, \citenamefont
  {Novak},\ and\ \citenamefont {Oertel}}]{Chatterjee:2016szk}%
  \BibitemOpen
  \bibfield  {author} {\bibinfo {author} {\bibfnamefont {D.}~\bibnamefont
  {Chatterjee}}, \bibinfo {author} {\bibfnamefont {A.~F.}\ \bibnamefont
  {Fantina}}, \bibinfo {author} {\bibfnamefont {N.}~\bibnamefont {Chamel}},
  \bibinfo {author} {\bibfnamefont {J.}~\bibnamefont {Novak}},\ and\ \bibinfo
  {author} {\bibfnamefont {M.}~\bibnamefont {Oertel}},\ }\bibfield  {title}
  {\bibinfo {title} {{On the maximum mass of magnetized white dwarfs}},\ }\href
  {https://doi.org/10.1093/mnras/stx781} {\bibfield  {journal} {\bibinfo
  {journal} {Mon. Not. Roy. Astron. Soc.}\ }\textbf {\bibinfo {volume} {469}},\
  \bibinfo {pages} {95} (\bibinfo {year} {2017})},\ \Eprint
  {https://arxiv.org/abs/1610.03987} {arXiv:1610.03987 [astro-ph.SR]}
  \BibitemShut {NoStop}%
\bibitem [{\citenamefont {Terrero}\ \emph {et~al.}(2019)\citenamefont
  {Terrero}, \citenamefont {Mederos}, \citenamefont {Pérez}, \citenamefont
  {Paret}, \citenamefont {Martínez},\ and\ \citenamefont
  {Angulo}}]{Terrero2019}%
  \BibitemOpen
  \bibfield  {author} {\bibinfo {author} {\bibfnamefont {D.~A.}\ \bibnamefont
  {Terrero}}, \bibinfo {author} {\bibfnamefont {V.~H.}\ \bibnamefont
  {Mederos}}, \bibinfo {author} {\bibfnamefont {S.~L.}\ \bibnamefont {Pérez}},
  \bibinfo {author} {\bibfnamefont {D.~M.}\ \bibnamefont {Paret}}, \bibinfo
  {author} {\bibfnamefont {A.~P.}\ \bibnamefont {Martínez}},\ and\ \bibinfo
  {author} {\bibfnamefont {G.~Q.}\ \bibnamefont {Angulo}},\ }\bibfield  {title}
  {\bibinfo {title} {Modeling anisotropic magnetized white dwarfs with gamma
  metric},\ }\bibfield  {journal} {\bibinfo  {journal} {Physical Review D}\
  }\textbf {\bibinfo {volume} {99}},\ \href
  {https://doi.org/10.1103/physrevd.99.023011} {10.1103/physrevd.99.023011}
  (\bibinfo {year} {2019})\BibitemShut {NoStop}%
\end{thebibliography}%

\end{document}